\documentclass[final,1p,times]{elsarticle} 
\usepackage{graphicx} 
\usepackage{amssymb} 
\usepackage{amsthm} 

\journal{Nuclear Physics A} 
\begin{document} 

\begin{frontmatter} 


\title{Elucidating Jet Energy Loss Using Jets: Prospects from ATLAS}

\author{N.~Grau$^{a}$, on behalf of the ATLAS Collaboration}

\address[a]{Columbia University, Nevis Laboratories, Irvington, NY 10533, USA}

\begin{abstract} 
Jets at the LHC are expected to provide the testing ground for studying QCD 
energy loss. In this contribution, we briefly outline the strategy that will 
be used to measure jets in ATLAS and how we will go about studying energy loss.
We describe the utility of measuring the jet $R_{AA}$, the fragmentation 
function, and heavy flavor jets. Utilizing the collision energy provided by 
the LHC and the nearly hermetic and highly segmented calorimeter, ATLAS is 
expected to make important contributions to the understanding of parton energy 
loss using fully reconstructed jets.
\end{abstract} 

\end{frontmatter} 



\section{Introduction}\label{sec:intro}
A wealth of measurements, interpreted as in terms of parton energy loss, have 
been made at the Relativistic Heavy Ion Collider (RHIC) in single particle 
measurements and multi-particle correlations~\cite{RHICCorr} in A+A 
collisions. 
Even at this time the debate continues on the mechanism or energy loss being 
either perturbative in nature, \textit{i.e.} gluon radiation, or some 
non-purturbative strong coupling phenomenon.
New experimental measurements are necessary in order 
to shed light on the mechanism and behavior of QCD energy loss. Interestingly, 
new results from full jet reconstruction in heavy ion collisions at RHIC are 
becoming available~\cite{RHICJetReco}. In this contribution, we present the 
prospects for measuring fully reconstructed jets using the ATLAS detector at 
the LHC. We outline several differential studies that can be performed with 
fully reconstructed jets and how they will extend our knowledge of what has 
been inferred about energy loss from measurements at RHIC.

The ATLAS detector is a large multi-purpose apparatus designed to measure rare 
high-$p_\mathrm{T}$ phenomena in high energy hadronic 
collisions~\cite{ATLASTDR}. Though designed for $p+p$ collisions, the 
capability of the detector for tracking, calorimetry, and muon identification 
are more than sufficient for use even in the extreme particle densities 
expected at the LHC~\cite{LoI}. For jet reconstruction, only the calorimeter 
is used in these studies. The ATLAS calorimeter consists of an electromagnetic 
calorimeter covering $|\eta|<3.2$ and full azimuth. The electromagnetic 
calorimeter consists of three sampling layers of varying radiation length and 
varying cell (single readout channel) sizes. The hadronic calorimeter, also 
consisting of multiple sampling layers, covers $|\eta|<5.0$ and full azimuth. 
Such a calorimeter presents unprecedented coverage for measuring jets in 
nuclear collisions at collider energies.

For this contribution, we focus on results from the cone algorithm for jet 
reconstruction. The ATLAS cone algorithm is a seeded cone algorithm run on 
calorimeter ``towers'', sums of EM and hadronic energy in bins of 
0.1$\times$0.1 in $\eta\times\phi$ (position) space. There is a requirement 
for one tower to exceed some predefined threshold seeding the algorithm. For 
each seed tower, the 4-vector sum of the towers within a cone in 
$\eta\times\phi$ space with radius $R$ is taken. The position of this 4-vector 
sum is then used as the new jet position. The 4-vector summing in a cone is 
iterated until the 4-vector sum position converges. It is possible that 
multiple towers in a jet are above the seed threshold producing multiple jets 
with overlapping constituents. An overlap fraction, the fraction of 
constituent towers shared between a pair of jets, is used to determine whether 
to combine or separate multiple jets, such that each tower belongs to one and 
only one jet. The ATLAS cone algorithm is defined by three independent 
parameters: 1) the cone radius $R$, 2) the seed threshold, and 3) the overlap 
fraction.

In heavy ion collisions, jet reconstruction must be modified to handle the 
underlying event energy. We have chosen to determine the underlying 
event energy, remove it from the calorimeter towers, and then run the cone 
algorithm as described above. The background determination is as follows. 
First, high-energy regions of the calorimeter are identified by performing a 
sliding window algorithm which calculates the scalar $E_\mathrm{T}$ sum of all 
3$\times$3 calorimeter tower regions. If a 3$\times$3 region has a summed 
$E_\mathrm{T}$ 2$\times$RMS above the mean of the summed $E_\mathrm{T}$ distribution, the 
central tower is tagged. The $\left<E_\mathrm{T}\right>$ of cells in each calorimeter 
sampling layer is computed for 0.1 unit bins in $\eta$ and by excluding cells 
that are within a cone radius of 0.8 units from any of the tagged towers. The 
$\eta$-dependence takes into account both physics modifying the underlying 
event distribution, as well as $\eta$-dependent material in ATLAS. The 
layer-dependence is important because some layers, especially the last 
longitudinal layers of the EM and Hadronic calorimeters, have little 
background to subtract. Tagging high energy regions in the calorimeter reduces 
the bias of jets on the underlying event $\left<E_\mathrm{T}\right>$ determination. 
Once the $\left<E_\mathrm{T}\right>$ is known, it is subtracted from \textit{all} of 
the cells of a given layer and $\eta$ bin. These subtracted cells are used to 
create new subtracted towers. These subtracted towers are then input to the 
jet reconstruction algorithm. This is, of course, only one method that could 
be used to handle the underlying event energy in cone jets. It should be noted 
that other algorithms are being studied, \textit{e.g.} $k_\mathrm{T}$ and anti-$k_\mathrm{T}$ 
algorithms. The $k_\mathrm{T}$ jet algorithm gives similar results to the 
cone~\cite{kTjets} while the anti-$k_\mathrm{T}$ study is ongoing. 

\section{Differential Jet Studies}\label{sec:jets}

\begin{figure}
\begin{center}
\includegraphics[width=0.45\linewidth]{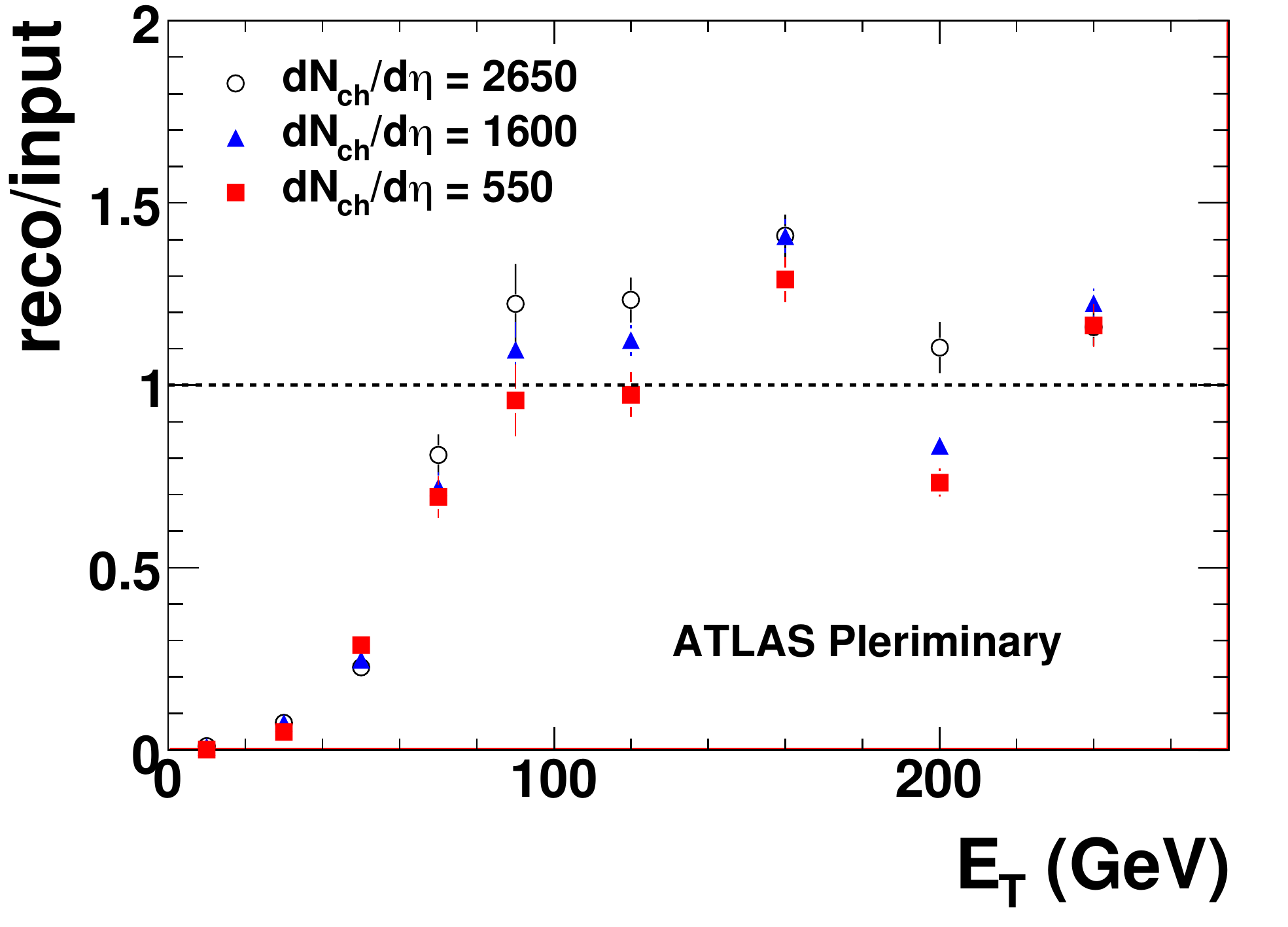}
\end{center}
\caption{Ratio of reconstructed to input jet spectrum for PYTHIA jets embedded 
in HIJING events. The jets are reconstructed with seed $E_\mathrm{T}$ = 10 GeV and $R$ 
= 0.4. There are no corrections for efficiency, energy scale, or energy 
resolution.}
\label{fig:jetraa}
\end{figure}

\begin{figure}
\begin{center}
\includegraphics[width=0.45\linewidth]{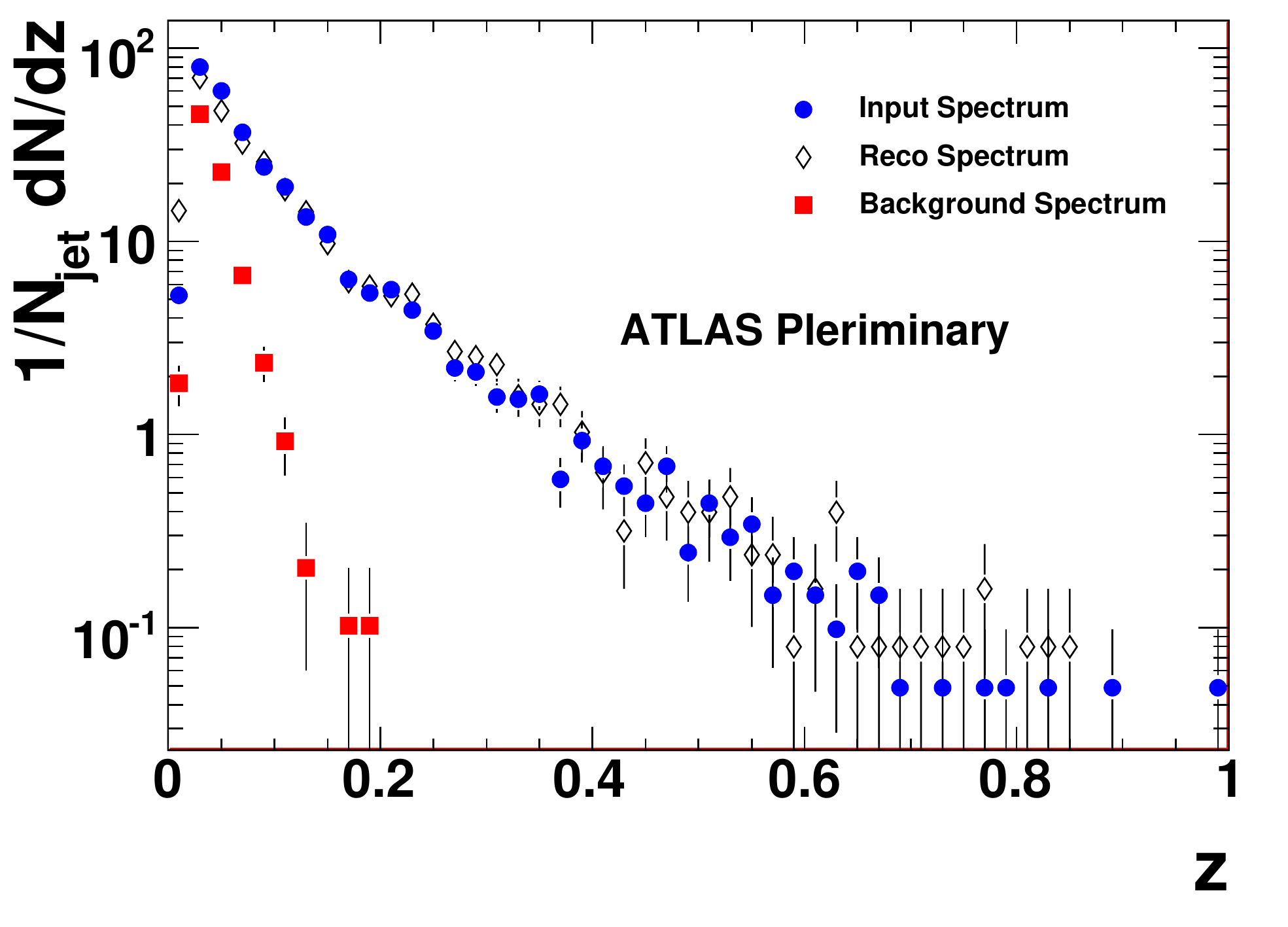}
\includegraphics[width=0.45\linewidth]{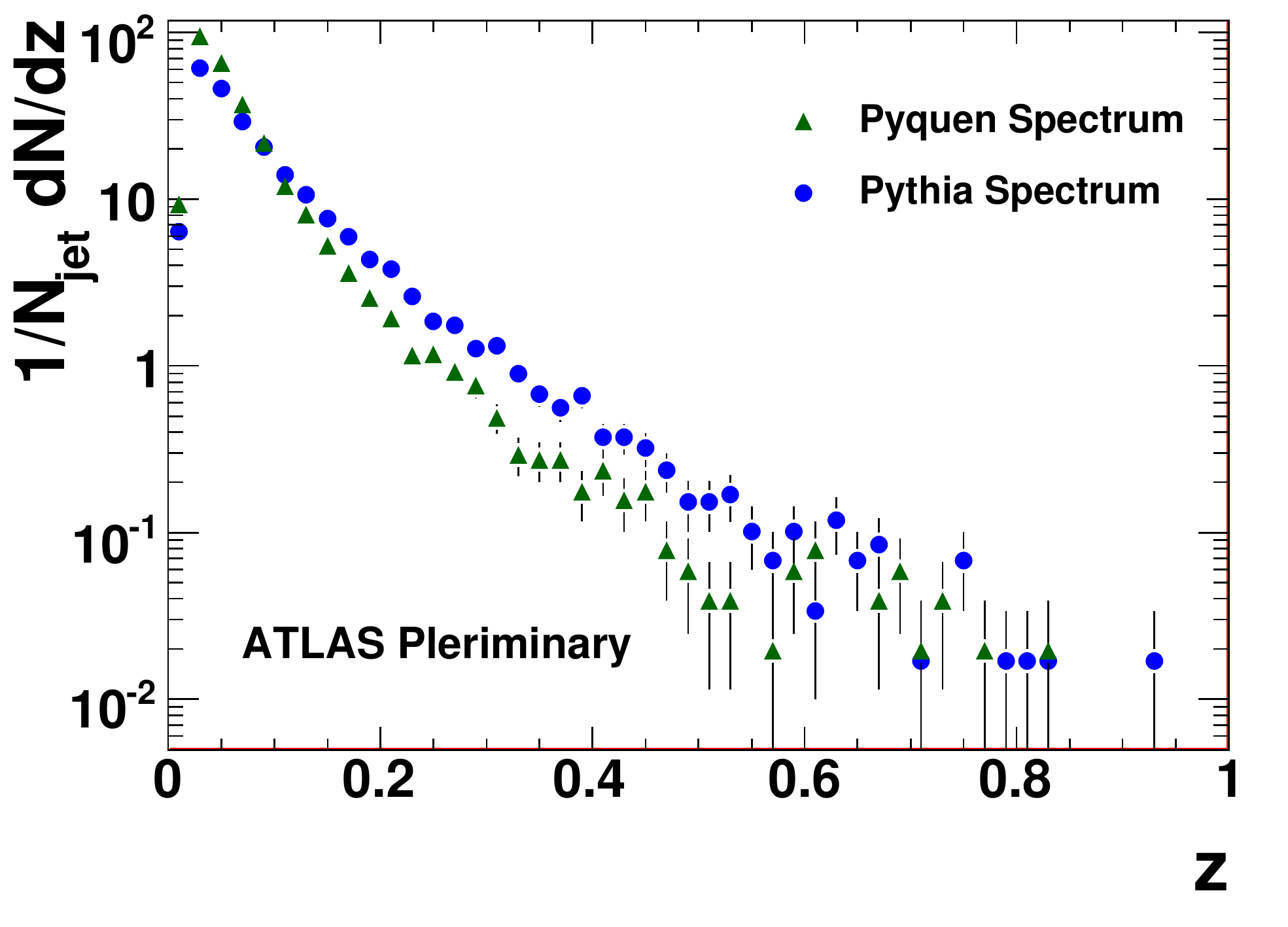}
\end{center}
\caption{\textit{Left:} The fragmentation function from input jets and input 
particles (circles), reconstructed jets and tracks (diamonds), and the fake 
distribution showing the contribution of the underlying event (squares). The 
fake distribution is already removed from the reconstructed spectrum. 
\textit{Right:} The fragmentation function from PYTHIA (circles) and PYQUEN 
(triangles) with the ATLAS detector response.}
\label{fig:jetdz}
\end{figure}

Once full jets are constructed, the goal is to further understand the 
mechanism of energy loss. One additional advantage of jet reconstruction, 
compared to single particle and di-hadron measurements, is the information 
gained by varying the parameters of the algorithms used. The measurements of 
jets in heavy ions can be performed as a function of not just centrality and 
reaction plane but also in the algorithm parameters.

One example of a differential study is measuring the jet $R_{AA}$ as a 
function of the cone radius $R$. According to a perturbative energy loss 
model~\cite{JetRaa}, cone jets reconstructed with different $R$ will be 
sensitive to different amounts of out-of-cone gluon radiation. In 
Ref.~\cite{JetRaa} the authors find that binary scaling of the jet 
cross-section is not recovered until $R=2$, which is probably difficult to 
measure experimentally. Still the variation of jet $R_{AA}$ as a function of 
$R$ should be sensitive to the radiated gluon angular distribution, a key 
prediction of perturbative energy loss models. Figure \ref{fig:jetraa} shows 
the ratio of reconstructed to input jet spectra as a function of jet $E_\mathrm{T}$. 
The reconstructed spectrum is not yet corrected for any reconstruction 
inefficiencies or resolutions. This shows that ATLAS will be sensitive to 10\% 
effects on the jet $R_{AA}$.

Another example is the fragmentation function, $D(z)$, which is the 
probability for a fragment to have a longitudinal energy fraction $z$ of the 
jet. The softening of the fragmentation function has long been an expected 
signature of parton energy loss since the leading, high-$z$, fragments are 
expected to lose energy, shifting down in $z$ while producing radiated gluons 
at low $z$~\cite{Armesto:2007dt}. In ATLAS fragmentation functions are 
constructed using charged tracks. These charged tracks are constructed in the 
inner detector with an efficiency of 70\% and fake rates $<$ 1\%. The tracks 
are then extrapolated to the calorimeter, and if the tower to which the track 
points is in a jet, it is considered a fragment. Only reconstructed tracks 
with $p_\mathrm{T} >$ 2 GeV are considered. The left panel of Figure \ref{fig:jetdz} 
shows the input PYTHIA fragmentation function compared to the fragmentation 
function after embedding into HIJING, full reconstruction, and application of 
a constant efficiency correction. The agreement between input and 
reconstructed spectra is quite good. This can be compared to the right 
panel of Figure \ref{fig:jetdz}, which shows the softer PYQUEN fragmentation 
function compared to PYTHIA even after simulating with the ATLAS detector 
response. We can expect to be sensitive to modifications at this level, a 
factor of 2 suppression for $z >$ 0.2, to the fragmentation functions.

\begin{figure}
\begin{center}
\includegraphics[width=0.36\linewidth]{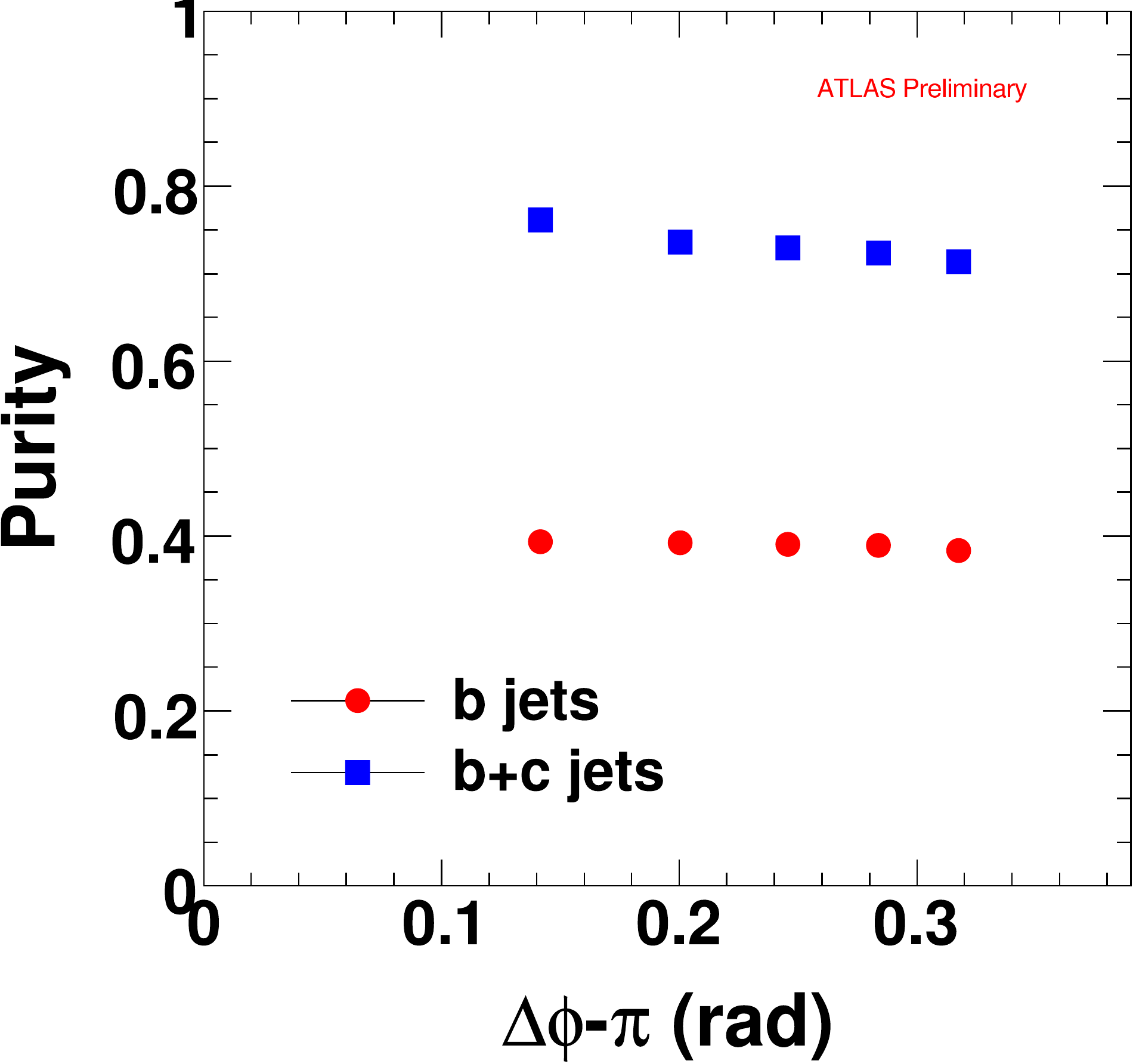}
\includegraphics[width=0.45\linewidth]{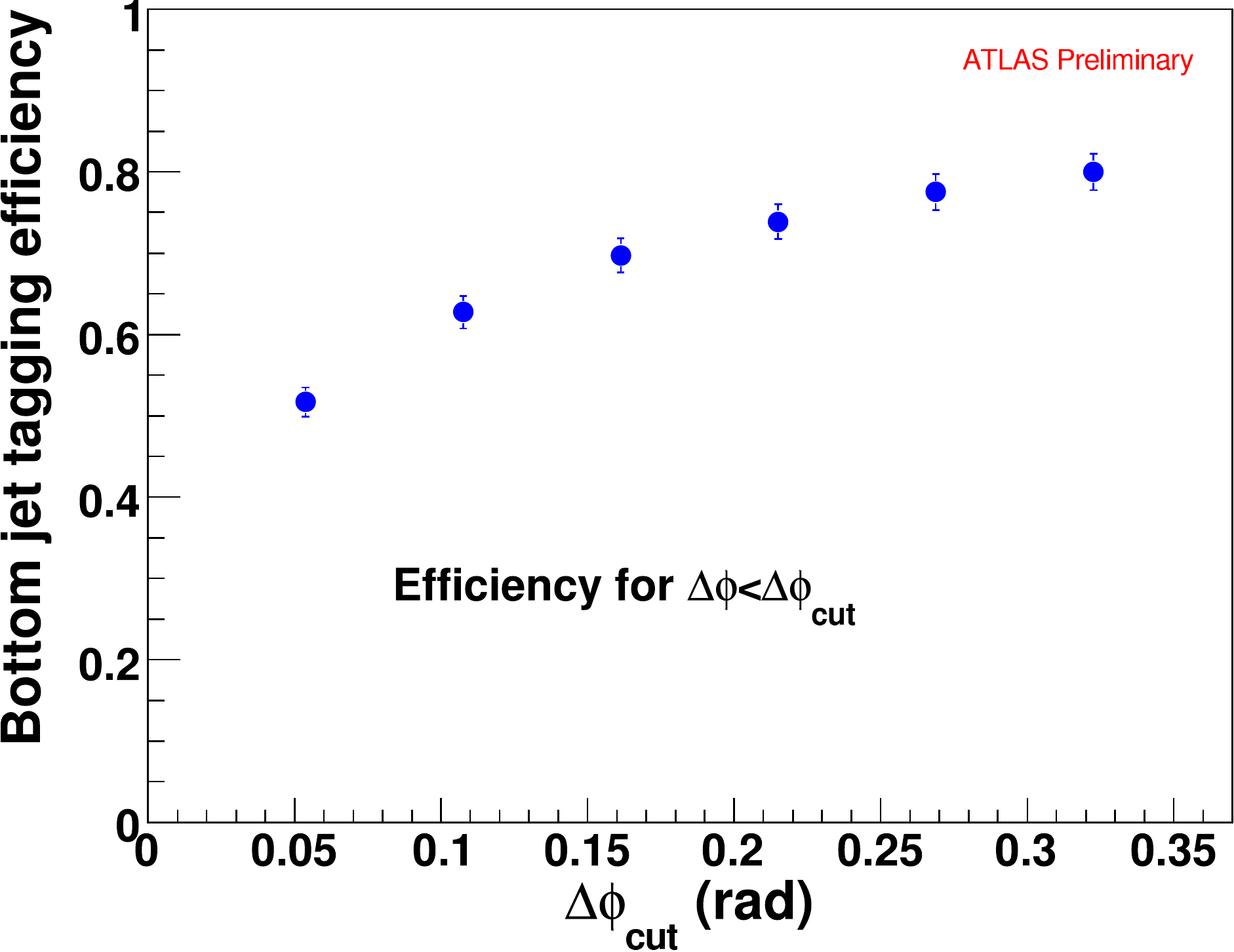}
\end{center}
\caption{\textit{Left:} Purity of tagging a bottom jet (circles) and heavy 
flavor jet (squares) requiring a $>$ 5 GeV muon associated with a jet below 
$\Delta\phi_{\mathrm{cut}}$. \textit{Right:} The efficiency of tagging a 
bottom jet with a given $\Delta\phi_{\mathrm{cut}}$.}
\label{fig:cbjets}
\end{figure}

Another topic of interest is heavy flavor energy loss. Early expectations from 
perturbative energy loss was that heavy flavor quarks lose less energy than 
light flavors. However, this expectation cannot be currently reconciled with 
single non-photonic electron data at RHIC~\cite{Adare:2006nq}. As a first 
attempt at understanding the ATLAS sensitivity to heavy flavor energy loss, we 
have studied our ability to tag heavy flavor jets with reconstructed muons. 
Full simulations of PYTHIA jets embedded into heavy ion events show a 
correlation in azimuth between heavy flavor (charm and bottom) jets above 70 
GeV and reconstructed muons above 5 GeV. The left panel of Figure 
\ref{fig:cbjets} shows the purity for tagging a heavy flavor jet or bottom jet 
as a function of the azimuthal angle cut imposed. The right panel shows the 
efficiency for tagging bottom jets using the muon. For 
$\Delta\phi_{\mathrm{cut}} <$ 0.15 rad, the bottom jet tagging efficiency is 
70\% with a purity of 40\% for b jets. Further techniques, such as 
reconstructing a displaced vertex associated with the jet, can be used to 
improve these numbers.

\section{Summary}\label{sec:summary}
In this contribution we briefly outlined the program for full jet 
reconstruction in heavy ion collisions with the ATLAS detector. Full jet 
reconstruction not only opens up new measurements, such as the jet $R_{AA}$ 
and fragmentation functions $D(z)$, but it also provides new information, 
through the variation of the jet reconstruction parameters for differential 
studies of energy loss on both light and heavy flavor jets. With its large 
acceptance tracking, calorimetry, and muon spectrometer, ATLAS is poised to 
make important contributions to the understanding of parton energy loss 
through the study of jets in heavy ion collisions at the LHC.




\begin{thebibliography}{00} 

\bibitem{RHICCorr}
J.~Nagle, these proceedings \\
A.~Sickles, these proceedings
\bibitem{RHICJetReco}
S.~Salur, these proceedings
\bibitem{ATLASTDR} The ATLAS Experiment at the CERN Large Hadron Collider \\
G. Aad, \textit{et al.} (ATLAS Collaboration) JINST \textbf{3}, (2008) S08003
\bibitem{LoI} ATLAS Heavy Ion Physics Letter of Intent, CERN-LHCC-2004-009
\bibitem{kTjets}N.~Grau (ATLAS Collaboration) arXiv:0805.4656 [nucl-ex]
\bibitem{JetRaa} I.~Vitev, S.~Wicks, and B.~Zhang JHEP \textbf{11} (2008) 093
\bibitem{Armesto:2007dt} N.~Armesto \textit{et al.} JHEP \textbf{02} (2008) 048
\bibitem{Adare:2006nq} A.~Adare {\it et al.} (PHENIX Collaboration), Phys.\ 
Rev.\ Lett.\  {\bf 98}, 172301 (2007)
\end{thebibliography}
\end{document}